\documentstyle[psfig,conf_iap]{article}
\begin{document}
\heading{%
%
Modelling the optically dark side of high--redshift galaxies
%
} 
\par\medskip\noindent
\author{%
Bruno Guiderdoni$^{1}$
}
\address{%
Institut d'Astrophysique de Paris, CNRS, 98bis Boulevard Arago, 
       F--75014 Paris.
}

\begin{abstract}
The recent detection of the Cosmic Infrared Background in FIRAS and
DIRBE residuals, and the observations of IR/submm sources by the ISOPHOT
and SCUBA instruments have shed new light on the optically dark side
of galaxy formation. It turns out that our view on the deep universe has 
been so far biassed towards optically bright galaxies. We now know that
a significant fraction of galaxy/star formation in the universe is hidden 
by dust shrouds. In this paper, we introduce a new modelling of galaxy
formation and evolution that provides us with specific predictions in the 
IR/submm wavebands. These predictions are compared with the current status 
of the observations. Finally, the future all-sky and deep surveys with the 
PLANCK and FIRST missions are briefly described.
\end{abstract}
\section{Introduction}
The corrections
needed to account for dust extinction in high--redshift galaxies
are rather uncertain, and might induce an
upward revision of the star formation rates (SFR) 
deduced from UV/visible observations 
by factors of a few (\cite{Pe} \cite{Me}). 
Moreover, star formation
might be completely hidden in heavily--extinguished galaxies which are
missed at optical wavelengths.
The discovery of the Cosmic Infrared Background in the FIRAS residuals
between 200 $\mu$m and 2 mm
(\cite{P1} \cite{G1}, \cite{F}) and in the DIRBE residuals at 240 and 140 $\mu$m
(\cite{Ha}) is a strong evidence that a significant fraction
of star formation is hidden in dust shrouds. The level of the CIRB
is more than five times higher than the ``no--evolution'' prediction 
based on the IRAS luminosity function, and twice as much as the level
of the Cosmic Optical Background deduced from converging faint galaxy counts.

Recent observations at 850 $\mu$m with the SCUBA instrument of the JCMT
(\cite{S} \cite{Hu}) have unveiled a 
large population of dusty objects. At the other
end of the submm range, the FIRBACK program with ISOPHOT aboard ISO
has yielded a catalogue of several hundred sources at 175 $\mu$m
(\cite{P2}). The
number counts are much higher than the ``no--evolution'' predictions.
Preliminary attempts for identification seem to show that many of 
these objects are likely to be at $z>1$.

We hereafter briefly describe a new modelling of galaxy 
formation (\cite{G1} \cite{G2}), and show predictions for on--going and 
forthcoming IR/submm surveys.

\section{Dust spectra in a semi--analytic framework}
The IR/submm spectra of galaxies are computed in the following way:
(1) follow chemical evolution of the gas;
(2) implement extinction curves which depend 
on metallicity as in the Milky Way, the LMC and SMC;
(3) compute $\tau_\lambda$;
(4) assume the so--called ``slab'' geometry where the star and dust 
components are homogeneously mixed with equal height scales. 
(5) compute a spectral energy distribution by assuming a mix of 
various dust components. The contributions are fixed in order to reproduce the 
observational correlation of IRAS colours with total IR luminosity.
\begin{figure}
\centerline{\vbox{
\psfig{figure=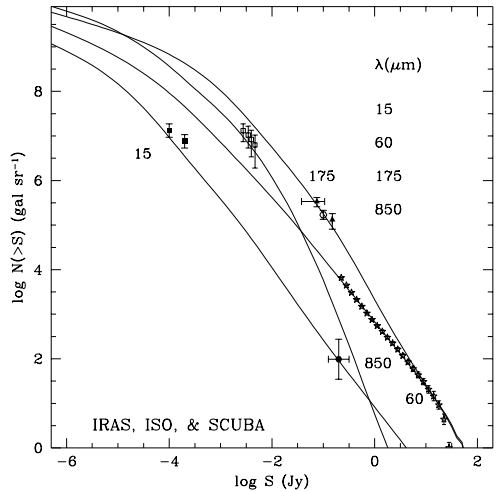,height=7.cm}
}}
\caption[]{Predictions for faint galaxy counts at 15 $\mu$m, 
60 $\mu$m, 175 $\mu$m, and 850 $\mu$m for a scenario fitting the CIRB,
and data. 15 $\mu$m: ISO--HDF follow--up with ISOCAM 
(\cite{O} solid squares),
IRAS survey (\cite{R} solid dot).
60 $\mu$m: IRAS survey (\cite{L} open stars). 
175 $\mu$m: ISOPHOT
Lockman Hole (\cite{K} open dot) 
and FIRBACK Marano field (\cite{P2} solid
triangles). 850 $\mu$m: JCMT/SCUBA (\cite{S} open squares). 
}
\end{figure}
These IR/submm spectra are implemented in a semi--analytic model of galaxy
formation and evolution. This type of model has been very effective in 
computing the optical properties of galaxies in the paradigm of hierarchical
clustering. We only extend this approach to the IR/submm range, and take
the standard CDM case with $H_0$=50, $\Omega_0=1$, $\Lambda=0$
and $\sigma_8=0.67$.
We assume a Star Formation Rate $SFR(t)=M_{gas}/t_*$, with $t_* \equiv \beta
t_{dyn}$. The efficiency parameter $1/\beta =0.01$ gives a nice fit of
local spirals. The robust result of this type of modelling is a rather flat 
cosmic star formation rate history. As a phenomenological way of
reproducing the steep rise of the cosmic SFR history from $z=0$ to $z=1$, 
we introduce a ``burst'' mode of star formation with ten times higher 
efficiencies.  This fairly reproduces the cosmic luminosity densities,
as well as the CIRB.

Fig. 1 gives the predictions of number counts at 15, 60, 175 and 850 $\mu$m
for the scenario which gives the best fit of the CIRB.
The agreement of the predictions with the data seems good enough 
to suggest that these counts do probe the evolving 
population contributing to the CIRB. The model
shows that 15 \%  and 60 \% of the CIRB respectively at 175 $\mu$m 
and 850 $\mu$m are built up by objects brighter than
the current limits of ISOPHOT and
SCUBA deep fields. The predicted median redshift of the ISO--HDF
is $z \sim 0.8$. It increases to $z \sim 1.5$ for the deep ISOPHOT
surveys, and to $z \ge 2$ for SCUBA, even if the latter value seems to be
very sensitive to the details of evolution.
\begin{figure}
\centerline{\vbox{
\psfig{figure=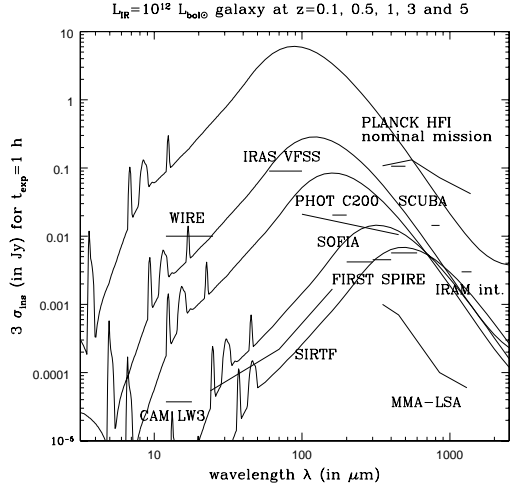,height=7.cm}
}}
\caption[]{An $L_{bol}=10^{12}$ $L_{bol\odot}$ galaxy, similar to 
Arp 220, at various redshifts, compared to instrumental sensitivities
for $t_{exp}=1$ h.
}
\end{figure}
\section{Future surveys with PLANCK and FIRST}
Fig. 2 gives a model spectral energy distribution typical of the 
$L_{bol}=10^{12}$  $L_{bol\odot}$ galaxy (similar to Arp 220) 
at various redshifts. 
The reader should note the specific behaviour of the observed flux at
submm wavelengths, where the shift of the 60 -- 100 $\mu$m 
rest--frame emission bump
counterbalances distance dimming. The instrumental sensitivities
of various on--going and forthcoming experiments are plotted on this
diagram: the IRAS  {\it Very Faint Source Survey},
ISOCAM, ISOPHOT, the IRAM interferometer, SCUBA, WIRE, SIRTF, SOFIA, 
the PLANCK {\it High Frequency 
Instrument}, the FIRST {\it Spectral and Photometric Imaging REceiver},
and the MMA/LSA.

The final sensitivity of these instruments is going to be confusion limited.
Table 1 gives the expected numbers of galaxies for the all--sky shallow
survey of PLANCK {\it HFI}, and the medium--deep survey of FIRST {\it SPIRE}
(to be launched by ESA in 2007), obtained from the model that fits 
the current counts.
Faint counts produced by these two instruments are complementary.
The study of the $250/350$ and $350/500$ colours are suited to point out 
sources which are likely to be at high redshifts. These sources can be 
eventually followed at 100 and 170 $\mu$m by the FIRST 
{\it Photoconductor Array Camera \& Spectrometer} and by the FTS mode
of {\it SPIRE}, to get the spectral energy distribution at $200 \leq \lambda 
\leq 600$ $\mu$m with a typical resolution $R\equiv \lambda /\Delta\lambda=20$.

\begin{center}
\begin{tabular}{l r r r}
\multicolumn{4}{l}{{\bf Table 1.} Predicted counts with PLANCK {\it
HFI} and FIRST {\it SPIRE}.}\\
\hline
\\
\multicolumn{1}{c}{$\lambda$}&\multicolumn{1}{c}{resolution}&
\multicolumn{1}{c}{$\sigma_{tot}$}&\multicolumn{1}{c}{$N(>5\sigma_{tot})$}\\
\multicolumn{1}{c}{}&\multicolumn{1}{c}{}&
\multicolumn{1}{c}{}&\multicolumn{1}{c}{}\\
\hline
\\
{\it HFI} 350 $\mu$m     & 5'  & 180 mJy   & 5400$^{*}$ \\
{\it HFI} 550 $\mu$m     & 5'  & 100 mJy   & 3800$^{*}$ \\
{\it HFI} 850 $\mu$m     & 5'  & 50 mJy    & 3100$^{*}$ \\
{\it SPIRE} 250 $\mu$m   & 18''&  3 mJy    & 70000$^{**}$ \\
{\it SPIRE} 350 $\mu$m   & 25''&  3.2 mJy  & 60000$^{**}$ \\
{\it SPIRE} 500 $\mu$m   & 36''&  4 mJy    & 27000$^{**}$ \\
\\
\hline
\end{tabular}

$^{*}$: 12 month nominal mission, 8 sr survey.\\ 
$^{**}$: 3000 h medium--deep survey, 900 sec/pointing, 60 deg$^2$.
\end{center}
The FIRST satellite is ideally designed to probe the bulk of the 
60 -- 100 $\mu$m rest--frame emission of galaxies at 
$1 < z  < 4$, and to recover the early SFR history of the universe. 
Our knowledge of 
the optically dark side of high--redshift
galaxies should increase spectacularly in the forthcoming years.

\vfill
\end{document}